\documentstyle[prl,aps,preprint,tighten,floats,psfig]{revtex}
\begin{document}
\draft
\preprint{IASSNS-HEP-95/113, UPR-0690-T, NSF-ITP-96-002
%, hep-ph/9512...
}
\date{February 1995}
\title{New Gauge Bosons from
  String Models\thanks{This paper is a summary, with a more
  phenomenological emphasis and additional discussion,
  of the results in our earlier
  article {\it Implications of Abelian Extended Gauge Structures
  from String Models},
  hep-ph/9511378.}} 
\author{Mirjam Cveti\v c$^{3}$
\thanks{On sabbatic leave from the University of Pennsylvania.} 
and Paul Langacker$^{1,2}$
}
\address{$^1$\ Department of Physics and Astronomy \\ 
          University of Pennsylvania, Philadelphia PA 19104-6396,\\
           $^2$ \ Institute for Theoretical Physics,\\ 
          University of California, Santa Barbara, CA 93106-4030,\\
           $^3$\  Institute for Advanced Study\\
           Olden lane, Princeton, NJ 08540\\ }
\maketitle
\begin{abstract}

We address the mass ranges of new neutral gauge bosons 
and constraints on the accompanying exotic particles
as predicted by a class of  superstring models.
Under certain  assumptions about the supersymmetry breaking
parameters we show that breaking of an additional 
 $U(1)'$  symmetry is radiative when the appropriate Yukawa 
couplings of exotic particles  are of order one, analogous
to the  radiative breaking of the electro-weak
symmetry  in the supersymmetric standard model due to the large top-quark
Yukawa coupling. Such large Yukawa couplings occur for a large
class of  string models. The $Z'$  and exotic masses
are either of  ${\cal O}(M_Z)$,
%when the symmetry breaking is due to a single standard model singlet,
or of a scale intermediate between the string and electro-weak scales.
%when the  symmetry breaking is due to two or more mirror-like singlets.
In the former case,  $M_{Z'}={\cal O}(1$ TeV$)$ 
may be achievable without excessive fine tuning, and is within
 future experimental reach.
\end{abstract} 

\section{Introduction}
New  neutral gauge bosons $Z'$  are a feature of many models  addressing the
physics beyond the standard model (SM).
The phenomenology of possible heavy gauge bosons
has been extensively  studied in the past. 
There are stringent limits on their mass and mixings
from precision electro-weak experiments \cite{L,precision} and
from direct searches  \cite{direct}.
The existence of a $Z'$ affects precision electro-weak data
(a) because $Z-Z'$ mixing pushes the $Z$ mass below the standard
model expectation; (b) this expectation is itself modified
by mixing, because it depends on the weak angle, the value of which
is confused or distorted by the effects of mixing on other observables.
(c) Both the mixing and the heavy particle exchange lead as well to
other changes in the predictions for the various observables,
implying new terms in the
effective interactions relevant to each process and leading to
different
apparent values of the weak angle determined in different processes.
Thus, the limits from precision experiments
vary significantly from model to model
because of the different chiral couplings to the
ordinary fermions. Typically, the mass of a 
heavy $Z'$ must exceed $\sim$ 400 GeV 
%(primarily from the CDF and D$^0$ experiments), while 
and the $Z-Z'$ mixing
angle must be smaller than a few times $10^{-3}$ in those models
in which the $Z'$ couples significantly to charged leptons, including the
standard examples from grand unification  \cite{L}. Models with
suppressed couplings to charged leptons can tolerate much larger
mixings (several percent), with the dominant constraint coming
from the shift in the light $Z$ mass, as is described in the Appendix.
Such models may be motivated \cite{zhadron}
by the possible enhancement in $Z \rightarrow b \bar{b} $ decays suggested
by the LEP data \cite{LEP}.
The direct production limits
\cite{direct} are likewise very sensitive to the $Z'$ couplings as
well as to the number of open channels for the decays ({\it e.g.}, into
exotic particles or superpartners), but
are again typically in the 300-600 GeV range for grand unification
type models.
%(from the  Z pole  analysis at $e^+e^-$ colliders (LEP and SLC)) 
%experiments.
%A more detailed discussion of the constraints is given in the 
%Appendix.

The identification and diagnostic study of heavy gauge bosons  
at future colliders has been investigated in 
detail \cite{CG}.
It should be possible to discover a heavy $Z'$ with mass up to $\sim 5$ TeV
through its leptonic decay channels, $pp\rightarrow Z' \rightarrow
\ell^+\ell^-$, $\ell= e$ or $\mu$, at the LHC, while $\gamma-Z-Z'$
interference effects should be observable at a 600 GeV $e^+e^-$
collider (NLC) for $M_{Z'}$ up to $\sim2$ TeV.  If such a $Z'$ exists,
it should be possible to obtain useful diagnostics about its coupling
by forward-backward asymmetries, rapidity distributions, rare decays
such as $Z'\rightarrow W\ell\nu$, and associated productions with a $Z$,
$W$, or $\gamma$, for masses up to $\sim3$ TeV at the LHC, with the
LHC and NLC providing complementary information. (For a review see Ref. 
\cite{CG} and references therein.)

 There have also been studies of the present and future
constraints on possible exotic particles, such as
heavy non-sequential quarks, leptons, or  standard model singlets
\cite{exotic}.
For example, some models predict the existence of a heavy,
$SU(2)_L$-singlet, charge $-1/3$ quark, $D_L-D_R$, which could
be produced at a hadron collider by ordinary QCD processes and
decay by $D_L-b_L,s_L,d_L$ mixing into, {\it e.g.,} $cW$, $bZ$,
or $bH$, where $H$ is a neutral Higgs boson.
Typically, the $cW$,
$bZ$, and $bH$ decays occur in the ratio $\simeq 2:1:1$. Currently, 
$m_D > $ 85 GeV if it mixes mainly with $b$ \cite{exotic}.
Additionally, precision experiments (weak charged current, neutral
current, and flavor changing constaints) typically
imply\footnote{Heavy gauge bosons and exotic matter constraints
have been addressed together in \cite{nrl}.}
that the
mixing between $D_L$ and $d_L$ is less than $\sim 0.01$ \cite{exotic}. 
Similarly, some models imply the existence of new 
$SU(2)_L$-doublets of leptons,
$\left( \begin{array}{c} E^+ \\ E^0 \end{array} \right)_L$ and
$\left( \begin{array}{c} E^+ \\ E^0 \end{array} \right)_R$,
which can be produced by ordinary electro-weak
processes, with the lightest decaying by mixing with ordinary leptons.
In both examples, the exotic particles are vector, {\it i.e.},
both the $L$ and $R$ fields have the same standard model transformation
properties. Vector particles could actually have bare masses as far
as the standard model is concerned, but in most theories there
are additional symmetries which forbid these and require that 
the masses be generated by some form of spontaneous symmetry breaking.
New vector multiplets  (or supermultiplets)
could in principle have arbitrarily large
masses since they do not require $SU(2)_L \times U(1)$ breaking, and they
are the most commonly predicted new type of matter in many standard
model extensions.
%Additionally, on purely phenomenological grounds there is no particular
%reason for the mass scale of new bosons or matter
%to be in the window ({\it e.g.}, up to a few TeV) accessible
%to present or future experiments.

In models   with extended gauge symmetry, 
but which do not incorporate constraints of
such underlying dynamics as grand unification,  supersymmetry,
supergravity, or string theory,  the mass  and couplings 
of additional gauge bosons are
free parameters, and thus  their masses can be anywhere 
from about 1
TeV to the Planck scale $M_{Planck}$. In addition, the  masses and couplings 
of the additional exotic particles 
which usually accompany the $Z'$ are also  free parameters.
Thus,  such  models  lack a unique  predictive power for 
$Z'$  physics at future experiments. 

The situation is different for  models that incorporate constraints
of supersymmetry: supergravity theories and, in particular, 
those that are predicted from string theory.
String theory  models necessarily include gravitational interactions (as a
consistent part of the theory),
and supersymmetry is restored  at sufficiently large energy scales. 
In addition, for  each of the
  models  the light particle spectrum 
%  at the large energy scale, {\it i.e.}, string scale $M_{string}$,
% is known 
as well as their couplings are calculable, thus providing
a major advantage over models in which 
both the particle content and the couplings
are put in the theory by hand.

Unfortunately, there are two
major hurdles that prevent 
one from obtaining unique low energy predictions 
of string theory: (i)
 by now a very  large number of  string models have been 
constructed, and one does
 not have a first principle guidance to single one model over another; 
(ii) while these models  possess supersymmetry at large energy scales, 
 one does not presently know how to 
break supersymmetry without introducing new parameters. 
Both problems are believed to have an ultimate resolution
 in the non-perturbative string dynamics. 

In spite of the above unsolved problems one can take a less
ambitious attitude and consider only those string models which have
a potential to be realistic. 
Those are string models  with supersymmetry,
  the standard  model (SM) gauge group as a part of the
 gauge structure, and a particle content that includes 
 three SM families \cite{FIQS,NAHE,F} and  at least two SM
Higgs doublets, {\it i.e.},  the string vacua which 
have at least the ingredients of the MSSM.
A number of such string models 
have been constructed \cite{NAHE,DHVW,ABK,CHL}. Also,
 one  may parameterize  our ignorance  
 of supersymmetry breaking
 by  introducing  supersymmetry breaking  mass and trilinear
interaction terms. 
  The mass terms ultimately drive the electro-weak symmetry breaking,
and thus are of the order of the
electro-weak scale. The models which have
been constructed often contain additional $U(1)'$ symmetries and
additional exotic matter multiplets.
% , as parameters in the theory.

%A set of models of that type constitutes a starting point to address 
%specific phenomenological issues. Here, we would like to derive 
%the consequences  of   an  additional  $U(1)'$ symmetry. 
%In contrast, 
A class of string models with the features mentioned above  
 and an  additional $U(1)'$ symmetry 
provide a  testing ground  to address  the  following  phenomenological
issues:  (i) a scenario,  which specifies the scale of  $Z'$ 
 and (ii) the mass  and phenomenological implications of the exotic 
particles accompanying the new gauge boson.   
We have identified several distinct scenarios, each of which can be  
illustrated by a specific  string model. 

A  main  conclusion is that  a large class of string models
 not only predict the existence of additional
gauge bosons, but  often imply the masses of
the new gauge bosons and  the exotic particles which necessarily accompany
them to be in the electro-weak range.
Each specific model leads  to  calculable  predictions (which, however, depend 
on
the assumed  supersymmetry breaking mass terms) for the masses, couplings, and
mixing with the $Z$ of the new boson(s), as well as for the
masses and quantum numbers of the associated exotic matter. 
We would  thus like to advocate that, from the string point of view,
new gauge bosons and associated exotic matter are perhaps the next best
motivated 
new physics -- after the Higgs and supersymmetric particles -- to be
searched for at  future experiments.

The paper is organized as follows. In Section 2 we specify 
in more detail the features of the string models and their advantage over  SM
physics. In Section 3 we  give the specific scenarios for the $Z'$ masses
achievable without excessive fine-tuning of the  supersymmetry breaking
parameters and constraints on exotic particles accompanying the new gauge
boson(s).
 We show examples in which the additional $Z'$ mass is: (a) comparable to
that of the $Z$ (already excluded); (b) in the  300 GeV to 1 TeV range,
which may still be
barely allowed but easily within the range of future or present
colliders; (c) at an intermediate scale ({\it e.g.}, $10^{8}-10^{14}$ GeV).
It is argued that in case (b) it is difficult though not impossible to
satisfy existing constraints on $Z-Z'$ mixing, especially for lower
values of $M_{Z'}$, and that $Z'$ masses above 1 TeV are not
expected (given our assumptions) without excessive fine tuning.
 Conclusions are given 
in Section 4. The Appendix discusses the limits on $Z-Z'$ mixing.

The results presented in this paper provide 
a summary  of 
Ref. \cite{CLI}, but with a more
phenomenological emphasis.  The  more detailed
version  \cite{CLI}  also  provides additional technical details and
illustrations of  the  scenarios within  specific  string models.

\section{Features  of String Models}

%We would like to emphasize  the constraining features of the superstring
%models as opposed to the  models where the particle content and the couplings
%are put in by hand. 
 
 Let us first specify    
the generic features of  supersymmetric  string models with  
the  standard model (SM) gauge group  $SU(2)_L\times U(1)_Y\times SU(3)_C$,  
three ordinary families,  and at least two SM
doublets, {\it i.e.}, a set of models with at least 
a particle content of the minimal supersymmetric standard model (MSSM). 
In addition, from a set of models we  select only
 those with  $SU(3)_C$, $SU(2)_L$ and $U(1)_Y$  
 all embedded into the $SU(5)$ gauge group, since for
  other types of embedding  the normalization of  the $U(1)_Y$ gauge group
  coupling is different from the one leading to the
   gauge coupling unification in the MSSM model.

In general these models also contain a   set  of particles  which are singlets
of the SM gauge group, but  which transform
non-trivially under  an additional  non-Abelian ``shadow'' 
sector group, and a  number of  additional $U(1)$'s, one of them 
 usually anomalous.  
The shadow sector non-Abelian gauge group  may play a role in 
dynamical supersymmetry breaking.
In addition, there are typically a large number of additional matter 
multiplets,
which  transform non-trivially under  $U(1)$'s  and/or the  standard 
model  symmetry.
In general such models also lead  to
%The fact that at $M_{string}\sim g_U\times 5\times 10^{17}$ GeV, 
%the string scale, where the gauge couplings of
%different gauge factors are unified and equal to $g_U$,
%  the observable sector gauge 
%group is  not $SU(5)$, but the SM gauge group, 
%implies \cite{S} that the theory in general contains 
fractionally charged color singlets, which may have important  
phenomenological\cite{AADF} consequences.

The anomalous $U(1)$  group  of such models 
is broken at $M_{string}$ \cite{DKI,DSW,ADS} by 
  nonzero  vacuum 
  expectation values (VEV's) of certain 
  multiplets which preserve supersymmetry  and consistency of the theory 
  at the loop level of string
  theory.
  At the same time,   this mechanism ensures that 
  a number of additional non-anomalous $U(1)$'s 
  are broken and  a number of  additional multiplets
become massive. Thus,  the enhanced  gauge symmetry and the  
exotic particle content 
is in general drastically reduced. Nevertheless,
there are often one or more non-anomalous $U(1)$'s and associated exotic
matter that are left unbroken.
The physics associated with  these left over 
non-anomalous $U(1)'$  
symmetries is the subject of this paper.

One should, however, point out that the models discussed 
in general may not be consistent with all of the phenomenological constraints:
(i) the models could have  additional color triplets  
in the spectrum which could   mediate a too fast 
proton-decay\cite {INQ,FPD}, (ii) the detailed  mass spectrum  
of the ordinary   fermions\cite{FIQS,F}  may not be realistic,
(iii) a scenario  for the symmetry breaking   of additional $U(1)'$s 
may not be consistent with
phenomenological  constraints on the exotic multiplets,
such as gauge coupling unification \cite{BL}.
In addition, it is not clear how to implement the supersymmetry breaking
scenario; we parameterize it by introducing  supersymmetry breaking mass
and cubic terms.

%Here, we shall concentrate on  phenomenological 
%consequences  of an additional non-anomalous $U(1)'$ symmetry. 
%We  shall  address the  specific 
%  aspects of $U(1)'$ symmetry breaking.

\section{$U(1)'$ Symmetry Breaking Scenarios}

Within the models discussed in the previous Section the pattern of additional
$U(1)'$ symmetry breaking is very constraining.  First, the models possess
supersymmetry, which relates the  interactions of integer  (boson) 
and half-integer (fermion) 
spin particles,  {\it i.e.}, those between the particle and its superpartner; 
second, the  the type of particles  and their couplings are calculable.
With the assumptions that supersymmetry is parameterized by 
 supersymmetry breaking mass parameters (and cubic scalar interactions),
 and that dynamical effects in
the shadow sector do not break the $U(1)'$ symmetry, the 
$U(1)'$ symmetry breaking must take place 
via the Higgs mechanism,  {\it i.e.}, by giving non-zero vacuum expectation
values (VEV)  to spin-zero particles $S_i$, which are
singlets\footnote{The possibility of breaking the $U(1)'$ summetry
by the VEV's of Higgs doublets is discussed in \cite{CLI}. In that
case, it is difficult to achieve a hierarchy between the $Z'$ mass and
the $(W,Z)$ masses.} under the SM symmetry, but
carry  non-zero charges  under the $U(1)'$. Non-zero VEV's of $S_i$ 
therefore preserve the SM gauge
symmetry; however, they yield a mass term  for $Z'$, which is of 
the order of the VEV's of $S_i$.
The analogous Higgs  mechanism  for spontaneous symmetry breaking (SSB) of the
electro-weak symmetry  takes place when the $SU(2)_L$ doublet(s) 
acquire nonzero VEV('s), 
giving mass to the $Z$ and $W^{\pm}$ gauge bosons.

We assume that the soft supersymmetry breaking mass-square terms for the
scalar fields are  positive
at $M_{string} \sim 10^{17}-10^{18}$ GeV and of the  order of gravitino mass 
$m_{3/2}$, which is assumed
to be of ${\cal O}(1)$ TeV.   
The appearance of positive mass-square terms
at $M_{string}$ is the case in almost all models that have
been explored. Then 
the only way of achieving SSB is via a radiative  mechanism.
 Namely, since such SM  singlets  have positive (supersymmetry breaking)
 mass-square terms at large energies, these terms need to  
be driven negative at lower energies to ensure  a
global minimum of their potential  with nonzero VEV's for such fields. 

The
 magnitude of the couplings  and mass parameters depends on the  energy 
  scale at which they are measured.
 The renormalization group equations account for the
 loop-corrections to  the physical  couplings and mass parameters, and 
 track the dependence of these couplings on the energy scale.
In particular,  some supersymmetry breaking 
 mass-square parameters of the Higgs field(s) $S_i$ can become negative if 
 $S_i$ have a large   Yukawa coupling (of ${\cal O}(1)$) to fermions
 with appropriate   representations  under the  SM gauge group, {\it e.g.},
  SM doublets or triplets.
%Such a scenario can be achieved if
%there are large Yukawa couplings of  the $S_i$'s to other fields, {\it
%e.g.}, 
%to a sufficient number of other fields.
% to SM doublets or color triplets.
In this case the loop corrections due to such Yukawa couplings (see Fig.1) 
contribute to the  renormalization group equations for
 the running  of the  mass-square terms in such a way that the 
 mass-square  term for $S_i$  can be driven negative,  while that 
for the supersymmetric partners of the fermions
(that couple to $S_i$ via the Yukawa couplings) remain positive (see Fig. 
2).\footnote{For the explicit form of  the
 renormalization group equations,  applied to specific string models,
 see the  appendix of Ref. \cite{CLI}.  For a detailed discussion
  of renormalization group equations within the MSSM see Ref. \cite{mssmrad}.}
\footnote{An analogous scenario takes place in the radiative 
breaking of the
 electro-weak symmetry in the MSSM.  There the positive (supersymmetry
 breaking) mass-square terms of the Higgs electro-weak doublet which couples 
to
 the top quark is driven to a
 negative value  at low energies due to the large  top-quark Yukawa coupling 
(which
 determines the mass of the top quark $\sim 170$ GeV), while the mass-square
 terms for the stops (supersymmetric partner of  the  top quark) remain
 positive.} Thus, the  scale of $U(1)'$ symmetry breaking depends  on 
 both the  type of  SM singlets $S_i$ responsible for the $U(1)'$ 
 symmetry breaking  and on the Yukawa  couplings of 
 such multiplet(s) to other exotic particles.

\begin{figure}
\begin{center}
\psfig{figure=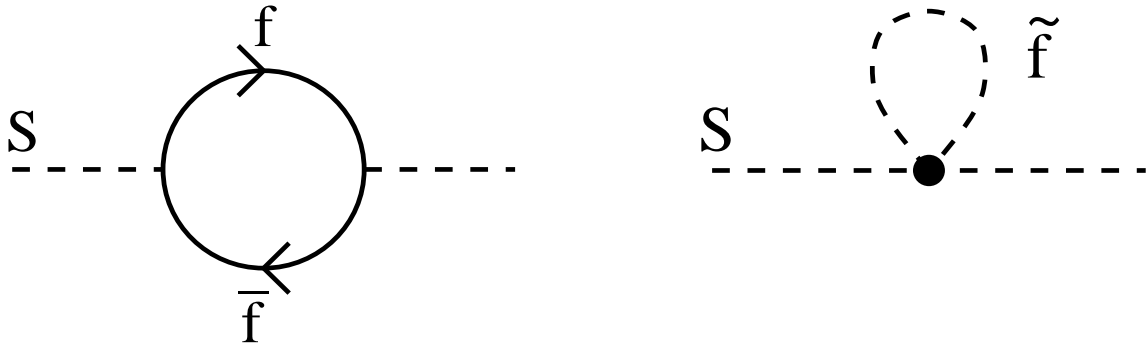,width=6in}
\end{center}
\vspace{1mm}
\caption{Loop corrections which lead to the
running of the  mass-square term for the scalar $S$. $f$ and $\tilde f$ denote the fermion(s) and their supersymmetric partner(s), respectively.}
\end{figure}

\begin{figure}
\psfig{figure=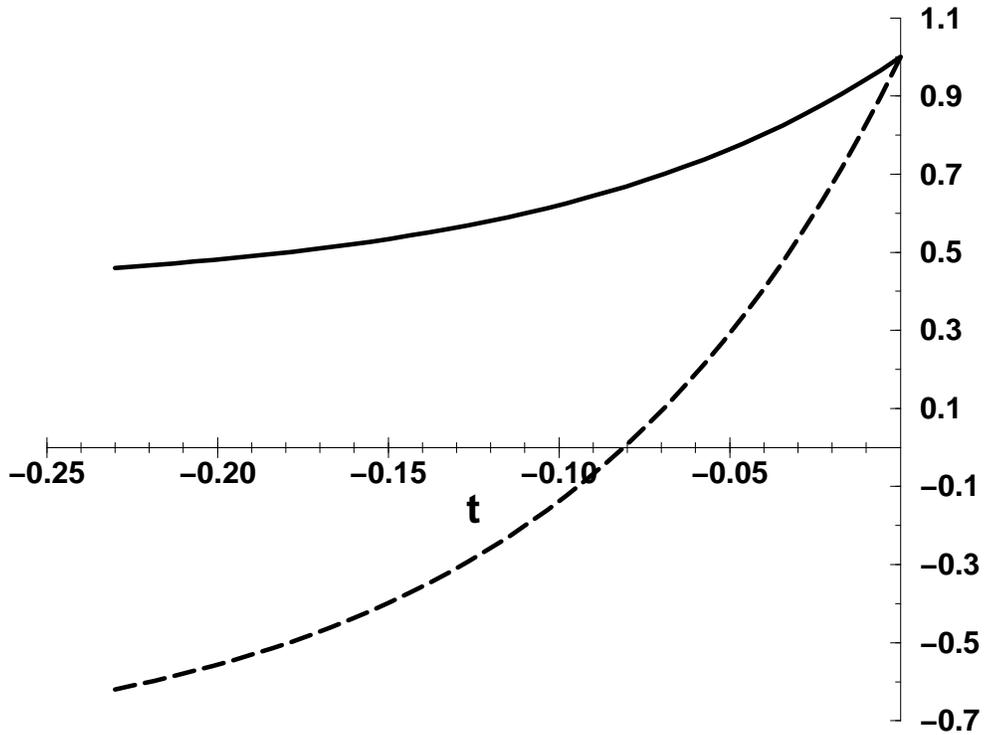,width=6in}
\vspace{1mm}
\caption[]{The running of the scalar mass-square terms  (in units of $m_{3/2}^2$) for a singlet field $S$
(dashes) coupled to a pair of $SU(3)$ triplets $D$ (solid line) (example 2 from the appendix
to \cite{CLI}). $t$, defined as $\log (\mu/M_{string})/16 \pi^2$, varies
from $-0.23$ at the electro-weak scale $\mu = M_Z$ to 0 at the string scale 
$\sim 5 \times 10^{17}$ GeV.}
\end{figure}

One of the features which distinguishes string models from the generic
MSSM is that the Yukawa couplings are calculable and thus are not  free 
parameters. Interestingly, 
for the  class of  string models
 discussed in the the previous Section
the corresponding Yukawa couplings are  either of 
${\cal O}(g)$   ($g$-gauge coupling) or zero\footnote{The same is true 
for the   Yukawa coupling of the ordinary families to the Higgs doublets.}.
Thus, if the relevant coupling is non-zero it  
may be sufficiently large to  ensure radiative breaking 
of the $U(1)'$ symmetry.

For  each of these possibilities the pattern of  $U(1)'$ symmetry breaking 
and  the running of the gauge couplings still depend on  the  specific 
exotic particle content and their couplings. 
We shall now discuss the possible scenarios for observable sector
$U(1)'$ symmetry breaking.
For the sake of simplicity  we shall address  scenarios in which
the electro-weak symmetry is broken  due to the non-zero VEV of 
the Higgs doublet  that couples to
the top-quark, {\it i.e.}, a large $\tan \beta$ scenario of the MSSM.
A generalization to scenarios that accommodate other ranges of $\tan \beta$
is straightforward. We will emphasize the general features which hold in each 
scenario.
However, we emphasize that in each specific model the $Z'$ mass, mixing,
and couplings, as well as the properties of the exotic matter, are
in principle calculable, though in practice they depend on the details
of the soft supersymmetry breaking.

\subsection{Symmetry Breaking Due to One  $U(1)'$ 
Charged Standard Model Singlet}

Suppose that the radiative
breaking of $U(1)'$ is due to one  SM singlet $S$, 
that is  charged under $U(1)'$. 
Namely, only  one $S$ has its effective mass-square  driven to a 
negative 
value at low energies, thus allowing for a non-zero VEV. 
The  $Z-Z'$ mass-square matrix is then:
\begin{equation}
M^2_{Z-Z'} =
\left ( \matrix{\frac{1}{2} G^2H^2 &Gg' Q'_HH^2  \cr 
Gg'Q'_H H^2& 2 g'^2({Q'}_H^2H^2+{Q'}_S^2S^2)} \right ),
\label{masscaseb}
\end{equation}
where  
$H$ and $S$ denote the  VEV's  for the SM Higgs doublet and singlet,
respectively, and $Q'_{H,S}$ are  the corresponding 
$U(1)'$ charges. Here  $G\equiv\sqrt{g^2+g_Y^2}$, where $g,g_Y,g'$ are the 
gauge couplings  at 
the SSB scale for  $SU(2)_L$,
 $U(1)_Y$ and $U(1)'$, respectively. 

The exotic matter to which $S$ couples acquires a mass of
order ${\cal  H}S$, where $\cal H$ is the relevant Yukawa coupling  between
the particular exotic  matter and  $S$.  In general, there
will be an additional  soft supersymmetry breaking mass term contributing to
the mass of the exotic matter even in the absence of the relevant Yukawa
coupling(s). 

The nature of the $Z-Z'$ hierarchy now crucially depends on
 the allowed VEV's $S$ and $H$, which are constrained by the
form of the  potential. At low energies
the potential can be written for the  particular direction with
non-zero VEV's as:
\begin{equation}
V=-|m_H|^2H^2-|m_S|^2S^2+{1\over 8}G^2H^4+{g'^2\over 2} (Q_H'H^2+Q_S'S^2)^2,
\label{potential}
\end{equation}  
where 
%we have assumed that $m_{H,S}^2>0$ (i.e., 
we have extracted
explicit minus signs from the negative mass-square terms.
%{\bf Our notation here (and in the first paper) is confusing. Sometimes
%we talk about running mass-square terms which can change sign, and
%sometimes we extract the explicit signs. Maybe we should introduce
%absolute values.}
Note that due to the constraints of
supersymmetry  the quartic couplings in (\ref{potential})  are proportional to
the  gauge couplings and  thus are {\it not}  free
parameters. 

One encounters the following two scenarios:
\begin{itemize}
\item (i): The relative signs of $Q'_H$ and $Q'_S$ are opposite.

 In this case the minimum of the 
potential  (\ref{potential}) is for:
\begin{equation}
H^2={{4(|m_H|^2+{{|Q'_H|}\over {|Q'_S|}}|m_S|^2)}\over {G^2}}\ \ ,\ \ 
S^2={{|m_S|^2}\over
{|Q_S'|^2g'^2}}+{{|Q'_H|H^2}\over {|Q'_S|}}
\label{VEVS}
\end{equation}
and the $Z-Z'$ mass-square matrix is
\begin{equation}
M^2_{Z-Z'} = 2
\left ( \matrix{(|m_H|^2+{{|Q'_H|}\over {|Q'_S|}}|m_S|^2) 
&{{2g'Q'_H }\over G} (|m_H|^2+{{|Q'_H|}\over {|Q'_S|}}|m_S|^2) \cr 
{{2g'Q'_H}\over G}
(|m_H|^2+{{|Q'_H|}\over {|Q'_S|}}|m_S|^2)& {{4g'^2{Q'}_H^2}\over{G^2}}
(1+{{|Q'_S|}\over{|Q'_H|}})(|m_H|^2+{{|Q'_H|}\over
{|Q'_S|}}|m_S|^2)+|m_S|^2 
} \right ).
\label{masscaseBa}
\end{equation}
It is difficult to achieve the needed
hierarchy between  $M_Z$ and $M_{Z'}$,
unless $|Q'_S| \gg |Q'_H|$ and $|m_S|^2 \gg |m_H|^2$, in such a way that
$|Q_H'/Q_S'|= {\cal O} (|m_H|^2/|m_S|^2)$.
The first condition is not normally expected to hold, except in the 
limiting case $Q_H' = 0$.

\item (ii):  The relative signs of $Q'_H$ and $Q'_S$ are  the same.

The minimum of the potential (\ref{potential}) now occurs for:
\begin{equation}
H^2={{4(|m_H|^2-{{|Q'_H|}\over {|Q'_S|}}|m_S|^2)}\over {G^2}}\ \ ,\ \ 
S^2={{|m_S|^2}\over
{|Q_S'|^2g'^2}}-{{|Q'_H|H^2}\over {|Q'_S|}}
\label{VEVSSAME}
\end{equation}
and the $Z-Z'$ mass-square matrix becomes:
\begin{equation}
M^2_{Z-Z'} =  2\left ( \matrix{(|m_H|^2-{{|Q'_H|}\over {|Q'_S|}}|m_S|^2) 
&{{2g' Q'_H} \over G}(|m_H|^2-{{|Q'_H|}\over {|Q'_S|}}|m_S|^2) \cr 
{{2g' Q'_H} \over G}(|m_H|^2-{{|Q'_H|}\over {|Q'_S|}}|m_S|^2)&
{{4g'^2{Q'}_H^2}\over{G^2}}
(1-{{|Q'_S|}\over{|Q'_H|}})(|m_H|^2-{{|Q'_H|}\over {|Q'_S|}}|m_S|^2)+|m_S|^2 
} \right ).\label{masscaseBb}
\end{equation}
In this case one encounters an interesting possibility for achieving a
hierarchy without an unusually small ratio of $ |Q_H'/Q_S'|$,
 provided $0<|m_H|^2-{{|Q'_H|}\over {|Q'_S|}}|m_S|^2
 \ll |m_S|^2$. 
 In this limit, the $Z-Z'$ mixing angle is
\begin{equation}
\theta_{Z-Z'} \sim {{2 g' Q'_H} \over{G}}{{M_Z^2} \over {M_{Z'}^2}}.
\label{thetath}
\end{equation}
For small $g' Q'_H/G$
the $Z-Z'$ mixing could be sufficiently suppressed to  be consistent
with the experimental bounds for  $M_Z'\le {\cal O}(1)$ TeV,
as is further discussed in the Appendix.

\end{itemize}

One can illustrate \cite{CLI} the above scenarios in a particular
class of string models.
The second case  is of interest, because there a reasonable hierarchy can be 
achieved  without 
fine tuning of the  soft supersymmetry breaking parameters or  choices of
models with unusual values of 
$|Q'_H|/Q'_S|$.  {\it E.g.,} such models can
provide for a  hierarchy 
$|m_H|^2-|m_S|^2|Q'_H|/|Q'_S|\ll |m_{S}|^2$, say,
$|m_H|^2-|m_S|^2|Q'_H|/|Q'_S|\sim |m_S|^2/10$. If, in addition, one has
{\it e.g.,} $g' |Q'_H|/G\sim 1/4$,
one  obtains
$M_Z'^2\sim 10 M_Z^2$ and the mixing angle $\theta_{Z-Z'}\sim  0.05$.
In this example, $M_{Z'}$
is barely within the current experimental bounds,
while $\theta_{Z-Z'}$ is too large for most choices of $Z'$
couplings \cite{L,direct}. Somewhat larger values of $M_{Z'}$ and
smaller values of  $g' |Q'_H|/G$ may be consistent with 
observations.\footnote{ In general, one also has to
ensure that the exotic particle content is compatible with the  unification of
the SM gauge coupling constants \cite{BL}.
This imposes another stringent constraint on
the  allowed exotic particle content.}
Thus, without excessive fine tuning of the
soft supersymmetry breaking parameters, the  prediction of $M_{Z'}$ is within
 experimental reach of present or future colliders. However,
when the  experimental bounds on $M_{Z'}$  exceed the 1 TeV region, this
scenario {\it cannot} be implemented without excessive fine tuning of the 
soft supersymmetry breaking parameters or unusual
choices of the  $U(1)'$ charge assignments.

\subsection{Symmetry Breaking Due to Mirror-like Pairs of $U(1)'$
Charged Standard Model Singlets}

\label{flatsection}
In this case, negative mass-square terms are induced for two (or more)
$U(1)'$ charged SM  singlets $S_{1,2}$, whose $Q_{S_1,S_2}'$ charges have
opposite sign. It turns out that supersymmetry constraints for the potential
ensure that the minimum of the potential  is along the VEV
 direction $Q'_{S_1}S_1^2=-Q'_{S_2}S_2^2\equiv S^2$  ($D$-flat direction). 
%which yields a potential  with no quartic terms.
 Along this direction the
 potential  for $S$  has no quartic terms; however, the mass square
 term depends on the energy  scale $\mu$ at which it is measured. Namely, 
one now has
to include the renormalization group improved potential, which  
 along the flat direction is of  the form :
\begin{equation}
V=  m_S^2(\mu=S)S^2
\label{flatpot}
\end{equation}
Thus, the minimum occurs near the  energy scale $\mu_{\rm crit}$ at which
$m_S^2$ turns negative.
In the case of radiative breaking with Yukawa couplings of ${\cal O}(1)$, 
it turns out that  $m_S^2({\mu_{\rm crit}})$ is
much larger than the soft supersymmetry breaking mass terms.  
For the models considered it is \cite{CLI} typically
four to ten
orders of magnitude below $M_{string}$. Therefore, in the case of flat
directions the  scale of symmetry breaking, {\it i.e.,} the VEV of $S$,
is  ${\cal O}(10^{-10}-10^{-4})
M_{string}={\cal O}(10^{8}-10^{14})$ GeV. 

Higher order (non-renormalizable) terms, which arise from exchanges of  massive
(of ${\cal O}(M_{planck})$) particles of string theory,  are also present in
the potential for $S$. They are of the type
$S^{2K+4}/M_{planck}^{2K}$ ($K\ge 1$) and 
could compete 
 with the radiative corrections included in (\ref{flatpot}). Such terms 
  determine  the scale  of symmetry breaking 
  to  be of the order of ${\cal O}([M_ZM_{planck}^K]^{1\over{K+1}})$. 
For example, for
$K=1$  the symmetry breaking scale is of the order $10^{11}$ GeV.

In either case the $Z'$  acquires mass 
of the order of the intermediate scale.
On the other hand, it is straightforward to show that the mass of the
physical Higgs boson associated with $S$ is of the order of
the soft supersymmetry breaking  mass terms. The exotic matter
 to which $S$ couples via the  Yukawa couplings  of magnitude ${\cal H}$  
acquires  mass of order ${\cal  H}S$,  {\it i.e.}, that of the intermediate
scale.   In the absence of the relevant Yukawa couplings, the
exotic matter would have a mass set
by  supersymmetry breaking mass terms, and  thus would be
of the weak scale.

Again there are string models which illustrate such scenarios \cite{CLI}, with 
the
appropriate
SM singlet  multiplets and the  necessary couplings to the exotic particles
to ensure the  symmetry breaking scenario discussed above.

Even in the absence of a $U(1)'$ gauge symmetry it is possible
for SM singlet scalar fields to acquire negative mass-square values at
low energies due to the radiative mechanism if they have sufficiently large
Yukawa couplings to other fields. 
Such scalars generally
will not have quartic terms in the potential, and thus they
would acquire intermediate scale VEV's.

\subsection{Other Implications of $U(1)'$}
Let us briefly comment on 
a few related topics. 

 We have seen that under a certain set of assumptions
the VEV's of standard model scalars will typically be either
at the electro-weak scale, or at an intermediate scale.
Intermediate scales are of interest in implementing the
seesaw model of neutrino mass. However, one may still need non-renormalizable
terms in the potential  to implement realistic neutrino
 mass scenarios \cite{CL}. Also, one
promising scenario for baryogenesis \cite{fy} is that a large
lepton asymmetry is generated by the decays of the heavy Majorana
neutrino associated with the seesaw, and then converted to a baryon
asymmetry during the electro-weak transition.
Such scenarios, while very attractive, cannot occur if 
there is a $U(1)'$ which is only broken at the electro-weak to TeV
scale \cite{buch}, unless, of course, the heavy Majorana neutrino is
a $U(1)'$ singlet.

We would also like to point out that  the string models with an additional
$U(1)'$  yield a natural solution the so called $\mu$ problem \cite{muprob}. 
Namely, the mass parameter $\mu$,  parameterizing  the 
 mixing the two SM Higgs doublets,  should be of the  order of the
 electro-weak   scale to ensure a realistic mass spectrum
 in the MSSM. In string theory, at large energy scales the  $\mu$ parameter
  is either of ${\cal O} (M_{string})$  or
 zero,  thus  causing  the phenomenological 
 $\mu$ problem.  If, in  the models with an additional $U(1)'$,  there is a
 coupling of the   the SM singlet(s)  $S$ to the two SM Higgs doublets and  $S$
 acquires a non-zero VEV's  of order 1 TeV, then such a term would provide  an
 effective $\mu$ term of the order of the electro-weak scale.

   We emphasize that 
the radiative  SSB scenarios discussed in this paper require the
existence of sufficiently large Yukawa couplings to drive the mass square 
values
of the SM singlet $S$ negative. This is most easily implemented if there
exists exotic matter which transforms non-trivially under the SM gauge
group. The exotic matter will then acquire mass terms given by the relevant 
Yukawa coupling times the VEV of $S$, as well as contributions from
 supersymmetry breaking mass terms.
Such exotic matter typically exists in string models.
It is usually vector, {\it i.e.}, both the left and right chiral
fields occur as $SU(2)$ doublets, or both as $SU(2)$ singlets.
However, if it carries SM quantum numbers it can destroy the
success of the gauge coupling unification \cite{BL}. Such effects largely
cancel if the light exotic matter corresponds to complete $SU(5)$
multiplets, but that is not typically expected in the types of
semi-realistic models we are discussing.
Preserving gauge unification without fine-tuning
is a stringent constraint on string model building, with or without
an additional $U(1)'$.

\section{Conclusions}

We have explored the possible scenarios
for (non-anomalous) $U(1)'$ symmetry breaking, as is expected
for a class of  string models with the 
standard model gauge group and additional $U(1)$ gauge factors
which survive into the low energy theory below $M_{string}$.
Under the assumptions that the additional $U(1)'$ symmetry  is not
broken by shadow sector effects and that
the  supersymmetry breaking scalar mass-square terms
are positive, the  breaking is necessarily radiative.
This requires  the existence of additional matter with large enough Yukawa
coupling to the standard model singlets responsible for the $U(1)'$ symmetry
breaking.  Then, within a particular model
 with definite   supersymmetry breaking mass  terms, the
 symmetry breaking pattern, the couplings and the masses of the
  new gauge bosons, and those of the
  accompanying exotic particles are  calculable.
In that sense the string models yield  predictions for the new physics
associated with the new gauge bosons.

It turns out that for the class of  string models considered the mass of $Z'$
is either of    ${\cal O}(M_Z)$  or  the 
intermediate scale of order  $10^{8-14}$ GeV. However, in both cases the
mass of the associated physical Higgs bosons is in the electro-weak region.
Our  major conclusion, therefore, is that  a large class of string models
considered here not only predict the existence of additional
gauge bosons and exotic matter, but   often imply that their masses should be
in  the electro-weak range.
Many such models are already excluded by indirect or direct constraints
on heavy $Z'$ bosons, and the $Z-Z'$ mixing is often too large, especially
for lower values of $M_{Z'}$.
The scenario in which  $M_{Z'}$ is in the
electro-weak range  allows, without excessive fine tuning of the
 supersymmetry breaking  mass parameters, for predictions of $M_{Z'}$ within
 experimental reach of present or future colliders.
On the other hand,  when the  experimental bounds on $M_{Z'}$  exceed the 1
TeV region, this
scenario {\it cannot} be implemented without excessive 
fine tuning of  supersymmetry breaking mass  parameters, or unusual
choices of  $U(1)'$ charge assignments.

String models provide  a concrete set of predictions for new gauge
boson  physics,
thus motivating  the $Z'$ searches   and   the 
associated exotic particle searches
at future experiments, as perhaps the best  motivated physics, next 
to that of the Higgs and
supersymmetric partner searches. 

%{\bf I must add appendix}

%{\bf References  need some work}
\acknowledgments
The work was supported in part by U.S. Department of Energy Grant No. 
DOE-EY-76-02-3071, the National Science Foundation Grant No. PHY94-07194, the
Institute for Advanced Study funds  and J. Seward
Johnson foundation (M.C.),
and the National Science  Foundation Career Advancement Award  
PHY95-12732 (M.C.).  We would like to thank G. Kane for encouraging us to write
a version of the paper  on 
new gauge bosons  with a phenomenological emphasis. We 
acknowledge hospitality of the Institute for Theoretical Physics, 
Santa Barbara, where the
 work was initiated.

\newpage

\appendix{\bf Appendix }

As described in the Introduction, the experimental constraints
on the $Z-Z'$ mixing angle $\theta_{Z-Z'}$ from precision
(mainly $Z$-pole) data are stringent but very model dependent.
It is common to quote limits on the masses and mixings of a
heavy $Z'$ with couplings given by a number of reference models.
These include models based on grand unification: the $Z_\chi$,
occurring in $SO(10)$ models, and the $Z_\psi$ and $Z_\eta$, which
correspond to two patterns of $E_6$ breaking. Other common 
reference models include the $Z_{LR}$, from left-right symmetric
models, and the $Z''$, which has the same couplings as the $Z$. The
latter would not occur in a realistic gauge theory, but is a 
convenient reference. The current 95\% CL limits on the mass and
mixing for each of these models are shown in the Table.
\begin{table}
\begin{tabular}{|l|ccccc|}
 & $Z_\chi$ & $Z_\psi$ & $Z_\eta$ & $Z_{LR}$ & $Z''$ \\
\hline
$\theta_{min}$  & -0.0041 & -0.0028 & -0.0069 & -0.0012 & -0.0027 \\
$\theta_{max}$  & 0.0011 & 0.0034 & 0.0020 & 0.0029 & 0.0004 \\
$M_{Z'}$ (precision) & 350 & 165 & 220  & 390 & 1000 \\
$M_{Z'}$ (direct) & 425 & 415 & 440 & 445 & 505 \\
\end{tabular}
\caption[]{95\% CL lower limit on $M_{Z'}$ in GeV and 95\% CL upper
and lower limits on the $Z-Z'$ mixing angle $\theta_{Z-Z'}$ for some
common reference models, from precision data (updated from \cite{L})
and direct searches \cite{direct,CG}. Recent (preliminary) CDF data
increase the direct  $Z''$ limit to 625 GeV.}
\end{table}

The limits in the Table assume that 
$\theta_{Z-Z'}$ and $M_{Z'}$ are independent parameters.
However, in specific models these are related if one knows
the $U(1)'$ charges of the Higgs doublets which break $SU(2)$
and generate the mixing. The relation is especially simple
if only one $SU(2)$ doublet is involved.
Assuming $M_{Z'} \gg M_Z$ the mixing angle
and masses are related by (\ref{thetath}). This relation
is more general than the string or grand unified models considered
here, and assumes only that there is a mass hierarachy and that
the mixing is dominated by a single Higgs doublet. In the reference
grand unified and left-right models one finds (assuming the
simplest Higgs structure and direct breaking of the grand unified
theory to the standard model with an additional $U(1)$) that
$2 g' |Q_H'|/G $ is in the range 0.2 -- 0.6, so the mixing
limits in the Table imply lower limits on $M_{Z'}$ around 1 TeV.
(In most cases the limits are relaxed when one allows two
or more Higgs doublets to contribute, since their contributions
can cancel in the off-diagonal term in the $Z-Z'$ mass matrix.)

For such couplings, the mass hierarchies that can be generated
without excessive fine tuning in the string models considered here
would be inconsistent  or only barely compatible with current
experimental limits. However,
the string-derived models are not expected to have
precisely the same couplings and $U(1)'$ charges as the
models based on simple grand unification, and one could have smaller
predicted mixings and/or less stringent experimental constraints.

For example, it is possible
that a new $Z'$ has suppressed couplings to charged leptons
or to all ordinary fermions.  Even if there are no selection rules
to make the relevant charges vanish, one expects $g' Q'$ to be
small if the $Z'$ couples to  a large numbers of exotic particles.
In a unified theory the overall strength of each gauge interaction, 
as normalized by $g'^2 \sum_i  Q_i'^2$,
should be the same or comparable.
This implies that if a $Z'$ couples to many particles, the
coupling to each particle individually is smaller, scaling
roughly as $1/\sqrt{N}$, where $N$ is the number of particles
to which it  couples. (This holds for the couplings
at the string scale. It applies to the running couplings as well if
the $N$ particles are all light.) In particular, if $g' Q_H'$
is small, then the predicted mixing is reduced. Similarly,
small $g' Q_e'$ leads to weaker experimental constraints on
$\theta_{Z-Z'}$, since, roughly, the combination 
$g' Q_e' \theta_{Z-Z'}$ enters most of the precision observables.
(Of course, suppressed couplings would also make the discovery of
a $Z'$ in the future more difficult).

As an extreme model, one can consider a $Z'$ that does not have any
couplings to ordinary fermions. The $Z-Z'$ mixing still affects
the precision observables because it shifts the $Z$ mass below
the standard model expectation (see below). A global fit to 
all data gives the best fit at $M_{Z'}$ = 130 GeV, although
the standard model ($M_{Z'}= \infty$) is allowed at 90\% CL, and
$\theta_{Z-Z'} = -0.05^{+0.03}_{-0.02}$.

To understand this relaxed constraint better, it is convenient to
consider the relation
\begin{equation}
\tan^2 \theta_{Z-Z'} = \frac{M_0^2-M_Z^2}{M_{Z'}^2-M_0^2}
\end{equation} 
between the mixing angle, the physical $Z$ and $Z'$ masses, 
and $M_0$, which is the the standard model prediction
for the $Z$ mass in the absence of mixing.
This constraint holds in all models, even when the $Z'$ coupling
to the ordinary fermions is absent. Assuming $M_Z \sim M_0$
and $M_{Z'} \gg M_Z$, this implies
\begin{equation}
| \theta_{Z-Z'} | \sim \frac{\sqrt{\rho_1-1}}{\sqrt{\rho_2^{-1}-1}},
\label{thetamz}
\end{equation}
where $\rho_1 = M_0^2/M_Z^2$ and $\rho_2 = M_0^2/M_{Z'}^2$.
The experimental error on $M_Z$ is negligible. However, the
standard model prediction for $M_0$ is uncertain
due to $\sin^2 \theta_W$, $m_t$, $M_H$, and $\alpha(M_Z)$,
which altogether yield an uncertainty
of $\sim \pm 0.03$ in $\sqrt{\rho_1-1}$. 
This uncertainty is increased to $\sim \pm 0.06$ if one allows
modest couplings of the $Z'$ to electrons, where we have assumed
that such couplings can distort the apparent $\sin^2 \theta_W$
obtained from $Z$-pole asymmetries by around 0.001.
In Figure 3 we indicate the range of upper limits on
$\theta_{Z-Z'}$ allowed by (\ref{thetamz}) for  $\sqrt{\rho_1-1}
= 0.03$ and $0.06$ as a function of $M_{Z'}/M_Z$. This is
the plausible experimental bound for models with suppressed
$Z'$ couplings to charged leptons.

The predicted mixing can be smaller
than the grand unification values for small $g' Q_H'$. It can
also be smaller
when two or more
Higgs doublets contribute significantly to the mixing,
as is expected to be the case in both superstring models and
supersymmetric grand unification ({\it i.e.}, away from the
large $\tan \beta$ limit). In this case, their contributions to
the mixing can cancel. A plausible (but not rigorous) range for the
effective value of $2 g' |Q_H'|/G $ is 0.5
to 0.1. This is also shown in Figure 3.

Thus, although many models are excluded by existing mixing constraints,
there is a reasonable amount of allowed parameter space, certainly
enough to motivate continued searches.

\begin{figure}
\psfig{figure=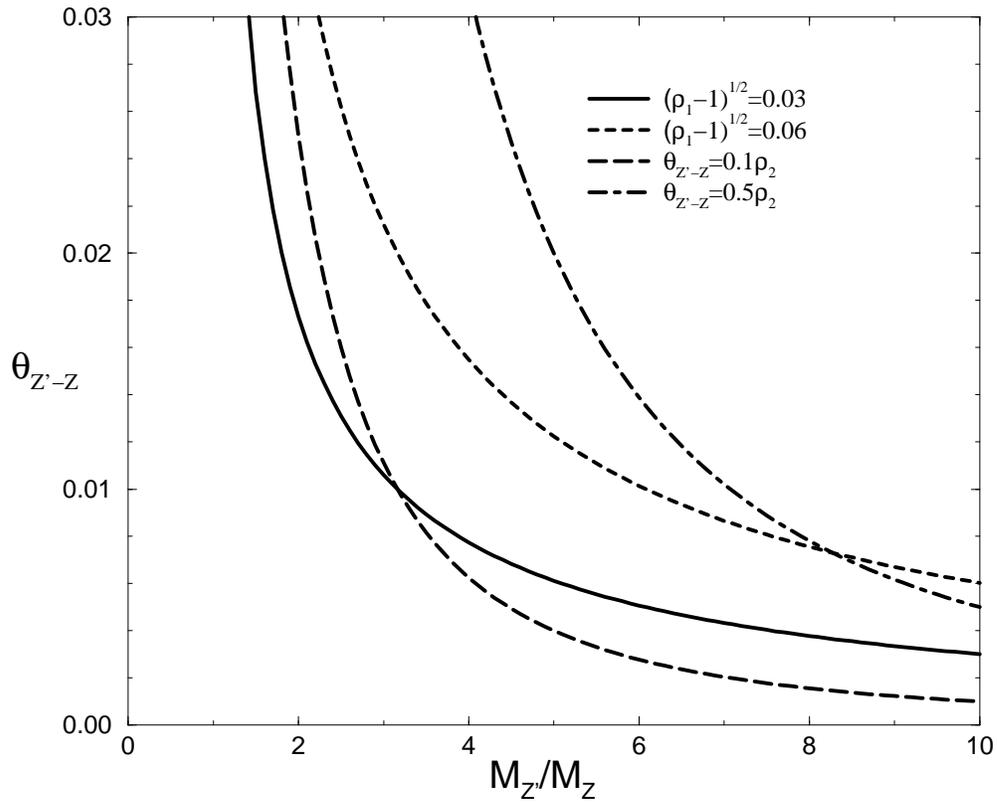,width=6in,angle=-90}
\vspace{1mm}
\caption[]{Theoretical expectation for $\theta_{Z-Z'}$ as a function
of $M_{Z'}/M_Z$ from (\ref{thetath}) for $2 g' |Q_H'|/G = 0.5$
and 0.1.  Also shown are the upper limits
on $\theta$ implied by the shift in the $Z$ from its standard
model expectation (eqn (\ref{thetamz})) for $\sqrt{\rho_1-1}=
0.03$ (experimental $\sin^2 \theta_W$ not affected by mixing), and 0.06
($\sin^2 \theta_W$ affected).}
\end{figure}
\end{document}